\documentclass[osajnl,twocolumn,showpacs,preprintnumbers,amsmath,amssymb]{revtex4}

\usepackage{graphicx}

\def\tauhz{\tau^{heat}_z}
\def\taucz{\tau^{cool}_z}
\def\tauhr{\tau^{heat}_y}
\def\taucr{\tau^{cool}_y}
\def\tauczr{\tau^{cool}_{z,y}}
\def\tauhzr{\tau^{heat}_{z,y}}

\begin{document}

\title{Doppler cooling of an optically dense cloud of trapped atoms}

\author{Piet O. Schmidt}
\author{Sven Hensler}
\author{J\"{o}rg Werner}
\author{Thomas Binhammer}
\author{Axel G\"{o}rlitz}
\author{Tilman Pfau}

\affiliation{5. Physikalisches Institut, Universit\"{a}t
Stuttgart,
  Pfaffenwaldring 57, D-70550 Stuttgart}

\begin{abstract}
  We have studied a general technique for laser cooling a cloud of
  polarized trapped atoms down to the Doppler temperature. A
  one-dimensional optical molasses using polarized light cools the
  axial motional degree of freedom of the atoms in the trap. Cooling
  of the radial degrees of freedom can be modelled by reabsorption of
  scattered photons in the optically dense cloud. We present
  experimental results for a cloud of chromium atoms in a magnetic
  trap. A simple model based on rate equations shows quantitative
  agreement with the experimental results. This scheme allows us to
  readily prepare a dense cloud of atoms in a magnetic trap with ideal
  starting conditions for evaporative cooling.
\end{abstract}

\maketitle 
\section{Introduction}
Since the advent of laser cooling of neutral atoms more than 20
years ago, this method has been extensively studied \cite{Lett:89}
and widely used for a variety of atoms and applications.
Ultimately, laser cooling followed by evaporative cooling
\cite{Ketterle1996c} allowed the creation of a Bose--Einstein
condensate (BEC) of neutral atoms
\cite{Anglin:2002,EFSummer:98,Cornell:2002}.

In almost all BEC experiments atoms are caught in a
magneto--optical trap. After sub-Doppler molasses cooling
\cite{Dalibard1989a} and polarization of the sample by optical
pumping, the atoms are typically transferred into a magnetic
 trap \cite{Anglin:2002,EFSummer:98,Cornell:2002} or, as has been
 recently demonstrated, into an optical
\cite{Barrett2001a} trap. The consecutive evaporative cooling
process requires both, a high initial density to provide a
reasonably large elastic collision rate, and a large number of
atoms, since most atoms are removed from the trap during
evaporation. The figure of merit for efficient evaporative cooling
is the initial phase space density of the atomic cloud. Therefore,
all laser cooling and polarization steps have to be optimized
carefully, to keep the temperature of the atoms low and the number
of atoms high, thus maximizing the phase space density in the
final trap.

In this paper, we present a robust Doppler cooling scheme in an
external trap which is applicable to most laser-coolable atomic
species.  Our scheme is particularly well suited for optically
dense samples and reduces the figure of merit for evaporative
cooling to the number of atoms transferred into the trap
regardless of temperature.

Free-space one-dimensional Doppler cooling is usually performed in
a standing wave light field created by two counterpropagating
laser beams with a frequency below the atomic transition
frequency. Cooling is based on preferential absorption of photons
from the laser beam opposing the direction of motion of the atom.
Subsequent spontaneous emission is centrally symmetric and does
not change the mean momentum of the atom. Doppler cooling of
polarized atoms in a magnetic trap has been proposed for atomic
hydrogen \cite{Hijmans1989a} and experimentally realized for
sodium \cite{Helmerson1992a}, hydrogen \cite{Setija:93} and
lithium \cite{Schreck:2001}. In Ref. \onlinecite{Helmerson1992a}
one-dimensional Doppler cooling of an optically thin cloud of
sodium atoms in a magnetic trap was performed. The atoms were
treated as a two level system since a very high magnetic offset
field ($B_0\approx 1500$~G) spectroscopically resolved the Zeeman
substates. Cooling of the motional degrees of freedom orthogonal
to the laser beams was provided by anharmonic mixing in the
trapping potential. Due to a long mixing time compared to the
cooling time, the achieved temperature was ten times the Doppler
temperature for sodium. In Ref. \onlinecite{Setija:93} single-beam
pulsed Doppler cooling was performed on a dense sample of
magnetically trapped hydrogen atoms. Thermalization of the cloud
was accomplished by elastic collisions between the atoms. Cooling
was limited to approx. five times the Doppler temperature due to
limited laser power and additional heating rates.

The main advantage of the cooling technique presented here is that
it combines three-dimensional temperatures close to the
Doppler-limit with the cooling of dense samples. In our
experiment, one-dimensional optical molasses cools the cloud of
atoms in axial direction down to the Doppler limit \cite{Lett:89}.
Cooling in the radial directions, orthogonal to the laser beams,
can be explained by reabsorption of spontaneously emitted photons
by the optically dense cloud. To our knowledge, reabsorption has
so far been treated only as a density-limiting mechanism which
comes along with heating of the atomic sample
\cite{Sesko1991a,Hillenbrand:1994,Ellinger:1997,Castin1998b}.
Here, we focus on the cooling aspects of scattered and reabsorbed
photons in a trapped polarized atomic sample.

We show that the atoms in the magnetic trap remain polarized
during the cooling process and practically no atoms are lost,
provided the magnetic substructure is spectrally resolved. We are
able to cool chromium atoms in a magnetic trap from $\approx 1$~mK
to a mean temperature of $240~\mu$K, corresponding to an increase
in phase space density by a factor of 80. We have studied the
dynamics of the cooling process as well as the steady state
temperature for various cooling parameters. The experimental
results can be explained with a simple model based on rate
equations for the temperature.

The paper is organized as follows. In Sec. 2, we present a
theoretical model for Doppler cooling in a magnetic trap in the
presence of reabsorption. The setup and experimental techniques
for our cooling experiments with chromium are described in Sec. 3.
Results on the dynamics and steady state properties of the cooling
process are presented in Sec. 4. We conclude with a discussion of
our results in Sec. 5.

\section{Theory}
\subsection{Rate Equations}
In this section, we present a simple model based on rate
equations. Although we assume cooling in a magnetic trap here, the
scheme is universal to all kinds of traps, provided cooling is
compatible with the trap and the quantization axis of the atoms is
independent of the position of the atoms in the trap.

Consider a cloud of spin-polarized atoms with mass $m$ and
magnetic moment $\mu$ confined in an axially symmetric harmonic
magnetic trap with trapping potential $V(x,y,z)=\mu
(\frac{1}{2}B_x^{''} x^2+\frac{1}{2}B_y^{''} y^2
+\frac{1}{2}B_z^{''} z^2 +B_0)$.

$B^{''}_{x,y,z}$ are the magnetic field curvatures in radial ($x$,
$y$) and axial ($z$) direction, respectively, and $B_0$ is a
homogeneous offset field along the axial direction. We assume that the
corresponding trap frequencies $\omega_{x,y,z}=\sqrt{\frac{\mu
    B^{''}_{x,y,z}}{m}}$ are much smaller than the Larmor precession
frequency at the center of the trap $\omega_L=\frac{\mu B_0}{\hbar}$, ensuring
that the atoms stay spin polarized and no spin flip losses occur
\cite{Sukumar1997a}.

Doppler cooling is performed on an electric dipole transition that
couples a long-lived ground state with total angular momentum $J$
to a short-lived excited state with angular momentum $J^{'}$, such
that $J^{'}=J+1$ \cite{Doppler_Foot1}. We also assume the
corresponding Land\'{e} factors $g_J$ and $g_{J^{'}}$ to have the
same sign, which we choose to be positive for our example. The
atoms are polarized in the "weak field seeking" state $m_J=J$ and
have a magnetic moment of $\mu=m_Jg_J\mu_B$, where $\mu_B$ is
Bohr's magneton.

Cooling is carried out in a $\sigma^+/\sigma^+$-standing wave created by two
laser beams propagating along the axial direction with normalized
intensity $I=I_\mathrm{laser}/I_\mathrm{sat}$ and detuning
$\Delta=(\omega_\mathrm{laser}-\omega_\mathrm{atom})/\Gamma$ from the unperturbed atomic
transition frequency. $I_\mathrm{sat}$ and $\Gamma$ are the saturation
intensity and the linewidth of the transition, respectively.

In general, the quality of the polarization of the cooling light
is important. For perfectly polarized $\sigma^{+}$-light
propagating along the quantization axis of the atoms in a
homogeneous magnetic field, no depolarizing transitions
$m_J\longrightarrow m_J, m_J-1$ can occur. This is also true for
excitation by scattered photons since the polarization with
respect to the quantization axis of the incident light is
preserved in the scattering process if the propagation direction
coincides with the orientation of the magnetic field. For
imperfect polarization, the cooling light can also drive
depolarizing transitions which cause loss of atoms. In the
configuration described above, the Zeeman-splitting due to the
finite offset field $B_{0}$ additionally reduces the scattering
rates $R_\mathrm{dep}$ of unwanted depolarizing transitions with
respect to the rate $R_\mathrm{pol}$ of the cooling transition
$m_J\to m_J+1$:
\begin{equation}
\label{rreq}
 R_\mathrm{dep}=\frac{\Gamma}{2}\frac{2IC_3^3}{1+4\Delta_{dep}^2}
\ll
\frac{\Gamma}{2}\frac{2IC_3^4}{1+4\Delta_{pol}^2}=R_\mathrm{pol},
\end{equation}
where $\Delta_\mathrm{pol}$ and $\Delta_\mathrm{dep}$ are the
effective detunings between the laser and the atomic transition
frequency in a magnetic field and $C_i^j$ is the square of the
Clebsch-Gordan coefficient for the transition from state $m_g=i$
to $m_e=j$.
\begin{figure}
\includegraphics[width=0.7\columnwidth]{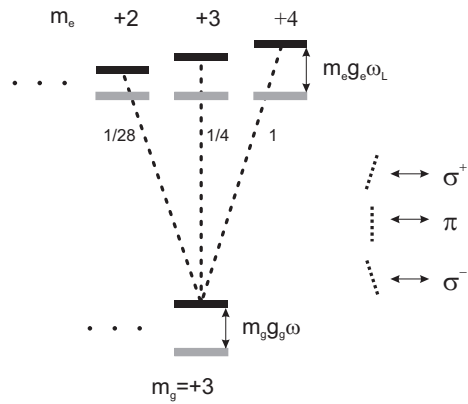}
\caption{\label{crzeeman}Part of the Zeeman substructure of
  $^{52}$Cr as an example of a $J\longrightarrow J+1$ transition.  Shown is the
  transition from the electronic ground state $^7S_3$ to $^7P_4$ with
  a wavelength of $\lambda_\mathrm{atom}=425.55$~nm, a saturation intensity
  $I_s=85.2$~W/m$^2$ and a linewidth of $\Gamma=2\pi\times 5.02$~MHz. The numbers
  next to the transitions are the squares of the Clebsch-Gordan
  coefficients.}
\end{figure}
In Fig. \ref{crzeeman} this effect is illustrated for the $J=3\longrightarrow J=4$
transition that connects the ground state $^7S_3$ to the excited state
$^7P_4$ in $^{52}$Cr. A full treatment of optical pumping in an
arbitrary light field taking into account all excited and ground state
levels gives the steady state population of each ground state for a
given magnetic field \cite{Stuhler:2002}. Assuming a polarization
quality of 1:10, a detuning of $\Delta=-0.5$ for the polarizing transition
and allowing for an atom loss of less than 1\% during the cooling
process, we find for the case of chromium a necessary offset field of
$B_0\geq 5$~G and for rubidium $B_0\geq 15$~G.

Following the common treatment of free-space Doppler cooling
\cite{Lett:89,Helmerson1992a}, rate equations for the temporal
evolution of the temperature for the axial ($z$) and radial ($y$)
degrees of freedom are readily obtained:
\begin{eqnarray}
\label{tratez}
    \frac{dT_z}{dt}&=&\frac{E_Rk_B}{2\tauhz}-\frac{T_z}{\taucz}\\
\label{trater}
    \frac{dT_y}{dt}&=&\frac{E_Rk_B}{2\tauhr}-\frac{T_y}{\taucr},\\
\end{eqnarray}
where $E_R=\frac{\hbar^2 k^2}{2m}$ is the recoil energy of the
cooling transition with wavenumber $k=2\pi/\lambda$ and $k_B$ is
the Boltzmann constant.  Usually, the heating and cooling time
constants are derived by considering the random walk in momentum
space due to spontaneous emission and the net cooling effect of
scattered photons from the two counterpropagating laser beams
\cite{Lett:89}. In an optically dense sample reabsorption of
photons has to be taken into account. Previously, reabsorption has
mainly been discussed as a reason for density limitations and
density-dependent heating in magneto-optical traps
\cite{Sesko1991a,Hillenbrand:1994,Ellinger:1997} and in the
context of sub-recoil cooling \cite{Castin1998b}. In our model, we
emphasize cooling effects of reabsorbed photons which can be
substantial for polarized atoms confined in an external trapping
potential.

The photon energy after a scattering event in the low intensity
limit \cite{Cohen:92doppler} is shifted by at most $4E_R+2\hbar
kv\ll\hbar|\Delta_\mathrm{pol}|\Gamma$, where $v$ is the velocity
of an atom at Doppler temperature. Therefore scattered photons
have essentially the same detuning and polarization as the cooling
laser beams. For our calculations, we use the axial direction
parallel to the laser beam as our quantization axis. The
contribution of the reabsorbed photons to cooling of the radial
and axial directions depends on the projection of their wave
vector $\vec{k}$ onto these directions. The effective intensity
$I^\mathrm{eff}_{z,y}$ of the cooling photons is proportional to
the incident laser light intensity $2I$ with a proportionality
factor $\kappa_{z,y}$ which primarily depends on the optical
density $OD_{z,y}$ in these directions:
\begin{equation}
\label{Ieff}
 I^\mathrm{eff}_{z,y}=2I\times\kappa_{z,y} \sim
2I\times \kappa^*_{z,y}\times OD_{z,y}
\end{equation}
If we consider a dipole radiation and absorption pattern for
$\sigma$-polarized light and take into account the $1/e^2$ size
$\sigma(\theta)$ of a Gaussian shaped cloud in the direction given
by $\theta$, we arrive at the following expressions:
\begin{eqnarray}
\label{kappar}
\kappa_y&=&\kappa_0\frac{2}{\pi}\int\limits_0^{\pi}d\theta(1+\cos^2{\theta})^2\sigma(\theta)\sin^2{\theta}\\
\label{kappaz}
\kappa_z&=&\kappa_0\int\limits_0^{\pi}d\theta(1+\cos^2{\theta})^2\sigma(\theta)\sin{\theta}|\cos{\theta}|
\end{eqnarray}
$\kappa_0=n_0\tilde{\sigma}_\lambda\frac{9}{128\sqrt{2\pi}}$ is an
angle-independent constant, with peak density $n_0$ and resonant
absorption cross section $\tilde{\sigma}_\lambda=6\pi\lambdabar^2$
which we assume to be uniform over the cloud. In a simple picture,
$\kappa_{z,y}$ can be interpreted as the effective number of
photons that contribute to cooling in the corresponding directions
originating from a single photon absorbed from the laser light.
The total number of reabsorption/emission cycles for an incident
photon is given by
\begin{equation}
  \label{kappa}
  \kappa=\kappa_0\int\limits_0^{\pi}d\theta(1+\cos^2{\theta})^2\sigma(\theta)\sin{\theta}.
\end{equation}
If the effects of reabsorbed photons are included the cooling
rates become \cite{Helmerson1992a}
\begin{eqnarray}
\label{taucz}
\frac{1}{\taucz}&=&-\Gamma\frac{32\Delta_\mathrm{pol}(1+\kappa_z)2I}{(1+4\Delta_\mathrm{pol}^2)^2}\frac{E_R}{\hbar\Gamma}\\
\label{taucr}
\frac{1}{\taucr}&=&-\Gamma\frac{32\Delta_\mathrm{pol}\kappa_y
2I}{(1+4\Delta_\mathrm{pol}^2)^2}\frac{E_R}{\hbar\Gamma}.
\end{eqnarray}
In axial and radial direction the total cooling light intensity
reads $I_z^\mathrm{tot}=2I+I_z^v=2I(1+\kappa_z)$ and
$I_y^\mathrm{tot}=I_y^v=2I\kappa_y$, respectively.

For the derivation of the heating rates, we have to take into
account directed reabsorption of $\kappa_{z,y}$ photons and
spontaneous emission of $(\kappa+1)$ photons for each photon
absorbed from the cooling laser beams:
\begin{eqnarray}
\frac{1}{\tauhz}&=&[(1+\kappa_z)+\frac{2}{5}(1+\kappa)]\frac{2I\Gamma}{1+4\Delta_\mathrm{pol}^2}\\
\frac{1}{\tauhr}&=&[\kappa_y+\frac{3}{10}(1+\kappa)]\frac{2I\Gamma}{1+4\Delta_\mathrm{pol}^2}
\end{eqnarray}
The factors $2/5$ and $3/10$ are statistical weights from the
dipole radiation pattern for the different directions of
spontaneous emission. In this model we have neglected additional
loss and heating processes, like e.g. radiative escape
\cite{Walker:1994} and radiation trapping \cite{Sesko1991a}. The
influence of these additional effects will be discussed in the
experimental part of this paper.

\subsection{Discussion of the Model}
\subsubsection{Steady State}
For simplicity we assume the coefficients $\kappa_{z,y}$ to be time
independent. This is justified for the evaluation of the steady state
parameters of the cooling process. The dynamics of the cooling process
can be described only qualitatively within this model, since the
optical density will change during cooling and therefore the number of
reabsorbed photons increases.

The steady state temperatures are readily derived from Eqns.
\ref{tratez}, \ref{trater} to be
\begin{eqnarray}
\label{Tzinf} T_z^\infty &=&\frac{\taucz}{\tauhz}\frac{E_Rk_B}{2}=\frac{1+\kappa_z+\frac{2}{5}(1+\kappa)}{1+\kappa_z}\frac{T_D}{2}\\
\label{Trinf} T_y^\infty
&=&\frac{\taucr}{\tauhr}\frac{E_Rk_B}{2}=\frac{\kappa_y+\frac{3}{10}(1+\kappa)}{\kappa_y}\frac{T_D}{2},
\end{eqnarray}
where $T_D=\frac{\hbar\Gamma}{2k_B}$ is the Doppler temperature.
These minimum temperatures are found at a detuning of
$\Delta_\mathrm{pol}=-\frac{1}{2}$, in agreement with free space
Doppler cooling \cite{Lett:89}. It is worthwhile to mention that
in this configuration in absence of reabsorption (i.e.
$\kappa_{z,y}=\kappa=0$) the 1D temperature in axial direction
\begin{equation}
    \label{Tmin}
    T_z^\mathrm{min}=\frac{7}{10}T_D
\end{equation}
is smaller than what is
usually called the 1D Doppler temperature. The reason is easily
seen: in the 1D model photons can be absorbed and emitted only
along the axial direction; in a 3D model with 1D cooling, photons
can be emitted into radial degrees of freedom according to the
dipol radiation pattern. These degrees of freedom take up photon
recoil heating and therefore reduce the heating rate in the axial
direction. In this situation the radial energy grows without
limit.

If we now include reabsorption, the situation changes. Whereas
reabsorption only in axial direction ($\kappa_z> 0$) changes the
cooling rate but not the steady state temperature, reabsorption in
radial direction ($\kappa_y> 0$) provides  a cooling effect, that
establishes a radial steady state temperature (Eq. \ref{Trinf}).
The magnitude of this temperature is determined by the ratio of
heating to cooling photons.

\begin{figure}
\includegraphics[width=\columnwidth]{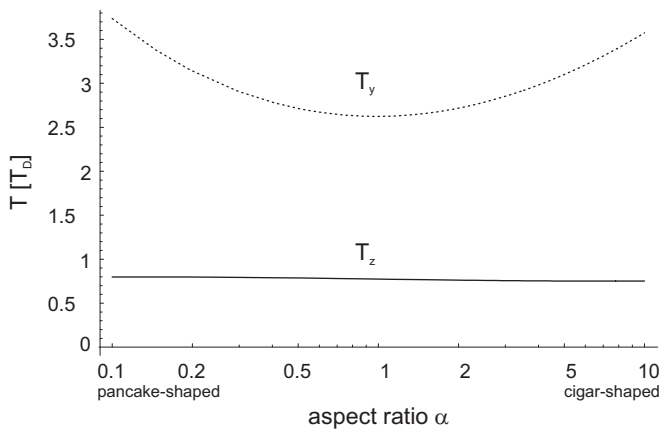}
\caption{\label{Tvsaspect}Temperatures in axial ($z$) and radial
  ($y$) direction as a function of trap aspect ratio for $N=10^8$
  atoms a peak density of $n_0=5\times 10^{10}~\textrm{cm}^{-3}$.}
\end{figure}
Experimentally, it is not possible to independently change
$\kappa_y$, $\kappa_z$ and $\kappa$. They depend via Eqns.
\ref{kappaz} and \ref{kappa} on the shape of the trapped cloud. In
Fig.\ \ref{Tvsaspect} we show an example, where we have integrated
these equations for different aspect ratios
$\alpha=\sigma_z/\sigma_y$ to calculate the steady state
temperatures. The aspect ratio has no influence on the axial
temperature, since cooling is dominated by photons scattered
directly from the laser beams. In radial direction we observe a
minimum near a spherical cloud. For a cigar shaped cloud the
optical density in radial direction decreases and for a
pancake-shaped cloud the solid angle decreases in which photons
contributing to radial cooling can be scattered.

\subsubsection{Dynamics}
The transient evolution of the temperature in radial and axial
direction can be approximated by the well known solution of Eqns.
\ref{tratez}, \ref{trater}:
\begin{eqnarray}
\label{Tzrt}
T_{z,y}(t)=T_{z,y}^\infty+(T_{z,y}^0-T_{z,y}^\infty)\exp(-\frac{t}{\tauczr})
\end{eqnarray}
where $T_{z,y}^0$ are the initial temperatures of the cloud in
axial and radial direction. Depending on the initial
temperatures and the steady state temperatures Eqns. \ref{Tzinf},
\ref{Trinf}, either Doppler cooling ($T_{z,y}^\infty<T_{z,y}^0$)
or Doppler heating ($T_{z,y}^\infty>T_{z,y}^0$) occurs. However,
since the atoms are oscillating in a harmonic trap, cooling
preferentially occurs close to the trap minimum where the velocity
of the atoms is highest. Therefore the cooling rates can not be
higher than approx. one quarter of the trap oscillation period
$t_{z,y}$.

\label{inttest}
According to Eqns. \ref{taucz}, \ref{taucr} the
time constants $\taucz$, $\taucr$ depend on the number of
scattered cooling photons. This can provide a test for the
reabsorption model. If cooling in radial direction originated from
elastic scattering \cite{Setija:93} or anharmonic mixing by the
trapping potential \cite{Helmerson1992a}, one would expect an
intensity independent relaxation towards an equilibrium
temperature, provided $\taucz\ll\taucr$, while our model explicitly
requires an intensity dependence of $\taucr$

We have already mentioned in the previous section that the
effective intensity of reabsorbed photons will change during the
cooling process. We will now discuss the dynamical effects on the
relevant parameters $\kappa_{z,y}$. From Fig. \ref{Tvsaspect} and
Eq. \ref{Tzinf} we deduce a weak dependence of the axial steady
state temperature on reabsorbed photons. In this regime cooling in
axial direction is dominated by the real laser beams and we can
assume $T_z$ to be unperturbed from its minimum value
$T_z^\mathrm{min}$. On the contrary, radial cooling strongly
depends on the reabsorbed photons, so we have to take into account
how $\kappa_y$ changes with time. As cooling in the $z$-direction
starts, the cloud shrinks according to
$\sigma_z=\sqrt{\frac{k_BT_z}{\mu B_z^{''}}}$. That will increase
the optical density in radial direction and thus the number of
reabsorbed photons according to Eq. \ref{Ieff} and
\[
\label{ODr} OD_y\propto\frac{1}{\sigma_z\sigma_y}\propto \kappa_y.
\]
As a result, the radial steady state temperature is reduced, since
it depends on the effective number of reabsorbed photons (Eq.
\ref{Trinf}). The same argument holds for the size of the cloud in
radial direction. During radial cooling $\sigma_y$ shrinks and we
arrive at an even lower $T_y^\infty$. In actual experiments this
nonlinear effect is less pronounced and will be neglected.
\section{Experimental Setup}
In our experiment, we continuously load typically $10^8$~chromium
atoms in the low field seeking $|j=3,m_J=3>$ magnetic substate into a
weak Ioffe-Pritchard trap \cite{Bergeman1987a} in the cloverleaf
configuration \cite{Mewes1996a}. Details of the continuous loading
mechanism can be found in Refs.  \onlinecite{Stuhler:2001,
  Schmidt:2002a}. In the compressed cigar-shaped trap with a nearly
harmonic trapping potential in all three dimensions, we have an
axial and a radial curvature of $B^{''}_z=110$~G/cm$^2$ and
$B^{''}_y=750$~G/cm$^2$, respectively.  This corresponds to
trapping frequencies for chromium $\mu=6\mu_B$ of
$\omega_x=\omega_y=2\pi\times 110$~Hz and $\omega_z=2\pi\times
42$~Hz. We use a rather high offset field of $B_0=28$~G to assure
sufficient harmonicity in all directions. During compression the
atoms are adiabatically heated to a temperature of 1~mK at a peak
density of $0.4-1\times 10^{10}$~cm$^{-3}$. In this trap we
perform one-dimensional Doppler cooling with a retroreflected
$\sigma^+$-polarized beam along the axial direction of the trap.
Laser light at 425.55~nm from a frequency doubled Ti:Sapphire
laser drives the $^7$S$_3 \longrightarrow ^7$P$_4$ transition in
$^{52}$Cr ($\Gamma=2\pi\times5.02~\mathrm{MHz}$,
$I_{sat}=85.2~\mathrm{W/m^2}$, see Fig. \ref{crzeeman}). The laser
frequency is tuned to $7~\Gamma$ above the unperturbed transition
frequency corresponding to an effective detuning of approx.
$-0.8~\Gamma$ for a nonmoving atom placed at the minimum of the
magnetic field in the center of the trap. This value was optimized
with respect to minimum temperature in all directions. The mean
intensity of the approx. 1~cm diameter beam as seen by the approx.
2~mm diameter cloud was measured to be $4\times 10^{-3}~I_{sat}$
using a 1~mm diameter pinhole in front of a calibrated photodiode
unless otherwise noted. The atomic ensemble is detected by
absorption imaging using a 12~Bit CCD camera. The resulting
optical density profile of the cloud is fitted with two orthogonal
Gaussian profiles meeting at the center of the cloud. The number
of atoms is extracted from this fit using the peak optical density
and the size.  We perform time-of-flight (TOF) sequences to obtain
the temperature in axial and radial direction. Densities are
derived using the size of the cloud in the trap obtained from the
time-of-flight fit and the number of atoms. The error bars shown
in the figures are the square root of the diagonal elements of the
covariance matrix for the fitting parameters obtained from a least
square fit. No systematic errors have been included, unless
otherwise noted.
\section{Experimental Results}
\subsection{Dynamics}
\begin{figure}
\includegraphics[width=\columnwidth]{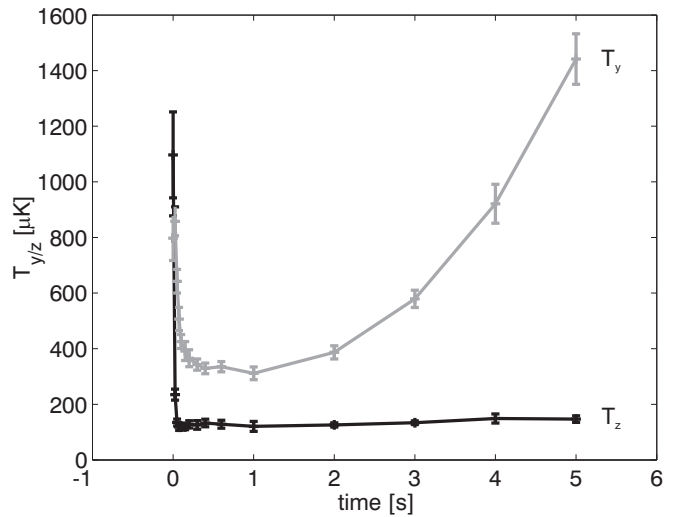}
\caption{\label{Tyztime}Typical evolution of the temperatures in
  axial ($z$) and radial ($y$) direction as a function of cooling
  time. The intensity of the cooling laser was $4\times 10^{-3}~I_{sat}$.}
\end{figure}
\begin{figure}
\includegraphics[width=\columnwidth]{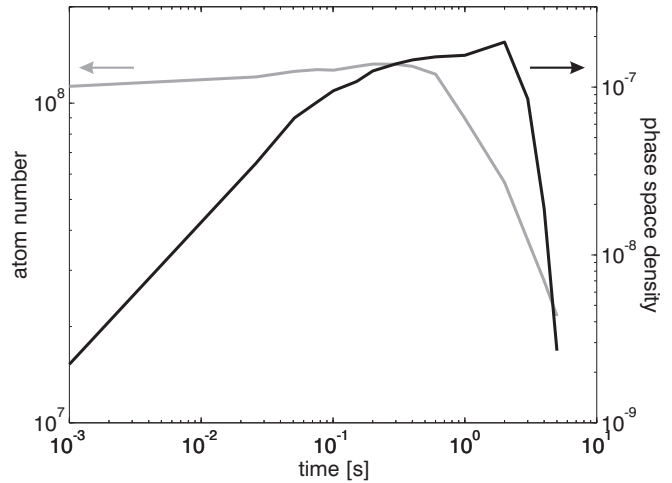}
\caption{\label{Npsdtime}Evolution of the
  number of atoms and the phase space density for the same data set as
  shown in Fig. \ref{Tyztime}.}
\end{figure}

In this section, we present experimental results on the dynamics
of the cooling process. In Fig. \ref{Tyztime} we have plotted the
temperature of the cloud in axial ($z$) and radial ($y$) direction
after a variable cooling time. We start in both directions with a
temperature of approx. 1~mK. In axial (cooling laser beam)
direction we see a fast decrease in temperature. But also in the
radial direction, where no laser beam is applied, cooling is
observed on a short timescale. The initial decrease in temperature
is very well fitted by an exponential decay in both cases, as is
expected from Eq. \ref{Tzrt}. The cooling time constants derived
from the fit are $\taucz\approx 10$~ms, corresponding to the
fastest possible cooling time of $\frac{2\pi}{4\omega_{z}}$ and
$\taucr\approx 50$~ms. In fact, we have never observed axial
cooling time constants significantly lower than 10~ms. The minimum
steady state temperatures in Fig. \ref{Tyztime} are
$T_y^\infty=334\pm 7~\mu$K and $T_z^\infty=124\pm 3~\mu$K,
corresponding to 2.7 and 1 times the 1D Doppler temperature of
$T_D=124\mu$K. Due to density dependent heating effects in the
light field \cite{Bell:99,Bradley:00a}, the minimum axial
temperature of $T_z^\mathrm{min}\approx 90\mu$K is not reached.
Fig. \ref{Npsdtime} shows the evolution of the number of atoms and
the phase space density during the cooling in a double logarithmic
plot to resolve the fast initial dynamics. The number of atoms
stays constant within experimental uncertainty beyond 300~ms at
which time the steady state temperature is reached. We gain more
than a factor of 80 in phase space density and increase at the
same time the peak density by more than one order of magnitude
from $0.4\times 10^{10}$ to $5.6\times 10^{10}~\mathrm{cm}^{-3}$.

In the experiment shown in Figs. \ref{Tyztime} and
\ref{Npsdtime}, we continued cooling after reaching steady state.
Fig. \ref{Npsdtime} shows a strong reduction in the number of
atoms for cooling times longer than 500~ms. Atoms are lost from
the trap presumably by radiative escape or fine structure changing
collisions \cite{Walker:1994, Stuhler:2001} in the cooling light.
At the same time the temperature increases in radial direction,
whereas it stays constant in axial direction. This is evidence for
a density dependent cooling effect in radial direction. A lower
number of atoms results in a reduced optical density and,
according to Eq. \ref{Trinf}, to a higher steady state temperature
$T_y^\infty$. We will come back to this effect in the discussion
of the steady state results.

\begin{figure}
\includegraphics[width=\columnwidth]{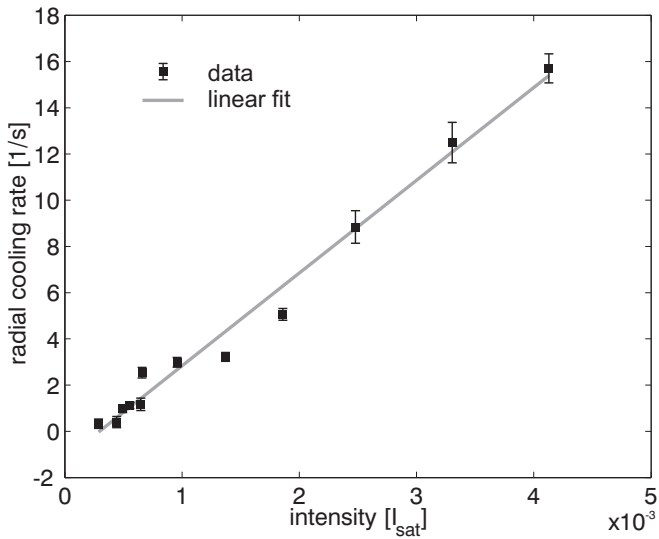}
\caption{\label{Ryint}Radial cooling rate $\frac{1}{\taucr}$ as a
function of measured single beam cooling light intensity. The
corresponding cooling rates in axial direction are larger by a
factor of at least 5 for each data point.}
\end{figure}
We have already pointed out in section 2\textit{B} that the cooling
rate of the radial degrees of freedom via anharmonic mixing or elastic
collisions between the atoms should be independent of cooling light
intensity, provided cooling in axial direction is much faster. In Fig.
\ref{Ryint} the results of a cooling experiment with different cooling
light intensities are plotted. For each data point we have performed
an experiment analogous to Fig. \ref{Tyztime} and fitted the initial
exponential decay of the temperature to obtain the cooling rate. The
corresponding rate constant for the axial direction has been verified
to be larger by at least a factor of five for each data point.  One
can clearly see a linear dependence of the radial cooling rate on
light intensity, ruling out anharmonic mixing or elastic collisions as
the major cooling mechanism. We have also performed an experiment in
which we stopped cooling after the axial direction reached steady
state, thus producing a highly anisotropic temperature distribution.
The subsequent cross-dimensional relaxation of the temperatures driven
by elastic ground state collisions occurred on a much longer timescale
than the observed cooling time constants when the light was on.

Increasing the laser intensity beyond the values shown in Fig.
\ref{Ryint} results in a strong distortion of the cloud due to light
pressure forces and intensity imbalance between the two
counterpropagating cooling beams. This causes parasitic heating and
increases the achievable temperatures.

From the slope of the linear fit to the data in Fig. \ref{Ryint}
we can extract $\kappa_y$ according to Eq. \ref{taucr}. Taking
into account that the average intensity incident on an atom is a
factor of 3 lower than the measured intensity due to absorption,
we obtain $\kappa_y=0.023$ for a detuning of
$\Delta_\mathrm{pol}=-0.8$. To compare this experimentally
determined value with theory, we have integrated Eq. \ref{kappar}
using the steady state size of the cloud $\sigma_z=700\mu$m and
$\sigma_y=410\mu$m. The theoretical result of $\kappa_y=0.1$ is
larger by more than a factor 4. This deviation supposedly arises
from uncertainties in the effective intensity as seen by the
atoms. This includes a systematic error in the measured intensity,
uncertainty in the averaged intensity due to absorption of photons
from the laser beam and averaging effects for the detuning due to
magnetic field curvature.

\subsection{Steady State Temperatures}
In this section, we present experiments that elucidate the
different effects influencing the steady state temperature and
compare them to theory.

\begin{figure}
\includegraphics[width=\columnwidth]{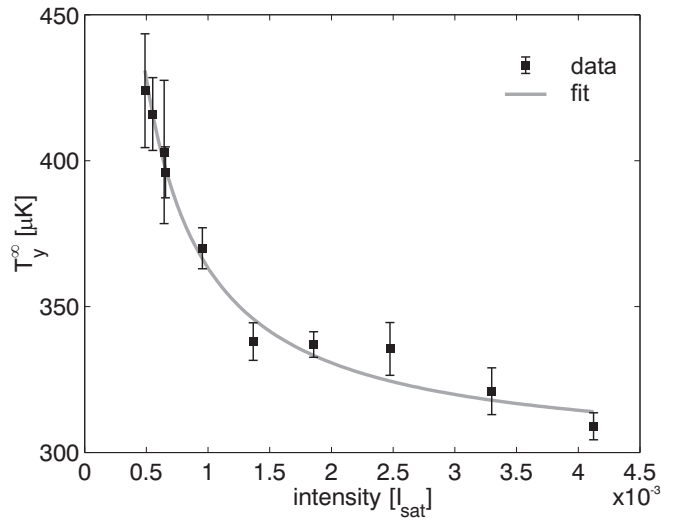}
\caption{\label{Tyint}Radial steady state temperature as a
function of cooling light intensity. The data is fitted by Eq.
\ref{TyinfR}. The data is from the same experimental run as in
Fig. \ref{Ryint}.}
\end{figure}
In the simplified model used to derive the steady state
temperatures in axial and radial direction given by Eqns.
\ref{Tzinf} and \ref{Trinf}, the latter are independent of cooling
light intensity, since both, the heating rate $1/\tauhzr$ and the
cooling rate $1/\tauczr$ are linear in intensity. Accounting for
an additional constant heating rate $R$ in the differential
equation (\ref{trater}), the steady state temperature for a
detuning of $\delta=-0.5$ becomes
\begin{equation}
\label{TyinfR}
T_y^\infty=\frac{\left(\kappa_y+\frac{3}{10}(1+\kappa)+\frac{R}{E_R\Gamma
I}\right)}{\kappa_y}\frac{T_D}{2}.
\end{equation}
In Fig. \ref{Tyint} we have plotted the measured steady state
temperature $T_y^\infty$ for different cooling light intensities.
The fit to Eq. \ref{TyinfR} using the measured detuning of
$\delta=-0.8$ and $\kappa=0.24$ calculated from Eq. \ref{kappa},
gives $\kappa_y=0.11\pm 0.002$ and $R=2.6\times 10^{-26}\pm
10\%~\mathrm{J s^{-1}}$. This result is in excellent agreement
with the integrated value for $\kappa_y=0.1$ obtained from Eq.
\ref{kappar}. From Fig. \ref{Tyint} we can also estimate a minimum
radial temperature of $T_y^\infty\approx 300 \mu$K which is in
good agreement with the predicted value using Eq. \ref{Trinf} and
Fig.  \ref{Tvsaspect} of $T_y^\infty=2.3 T_{D}\approx 290\mu$K.
Thus, the additional, intensity independent heating mechanism
slightly increases the minimum temperature. We have identified
this mechanism to be a strong dipolar relaxation process in
chromium \cite{Schmidt:2002b}.

\begin{figure}
\includegraphics[width=\columnwidth]{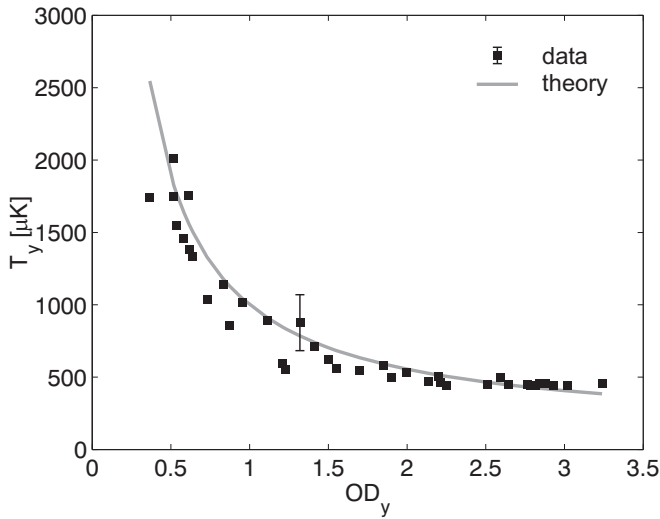}
\caption{\label{TyODy}Steady state temperature in radial direction
as a function of optical density (OD) in this direction. The error
bar resembles the typical error from the temperature fit to the
time of flight data. The theory curve is a plot of Eq.
\ref{TyinfRODy} using the results of the fit from Fig.
\ref{Tyint}.}
\end{figure}

We have already seen in Fig. \ref{Tyztime} that the steady state
temperature in radial direction depends on the number of atoms in
the cloud (increase in temperature for a decreasing number of
atoms at long cooling times). To elucidate this effect more
quantitatively we have performed an experiment in which we cool
down the atoms to steady state as described before. In a second
"cooling" stage we reduce the cooling laser intensity to
$1.6\times 10^{-3}~I_\mathrm{sat}$ to slow down the loss of atoms.
After different cooling times, we measure the temperature and the
optical density in radial direction. The result is plotted in Fig.
\ref{TyODy} as a function of optical density in radial direction.
We observe a decrease in temperature with increasing optical
density. The effective number of reabsorbed radial and total
cooling photons, $\kappa_y$ and $\kappa$, respectively, are to
lowest order linearly proportional to the optical density $OD_y$
(Eq. \ref{Ieff}). Introducing two new constants,
$\kappa_y^*=\kappa_y/OD_y$ and $\kappa^*=\kappa/OD_y$, the steady
state temperature reads
\begin{equation}
\label{TyinfRODy}
T_y^\infty =\frac{\left(\kappa_y^*\times
OD_y+\frac{3}{10}(1+\kappa^*\times OD_y)+\frac{R}{E_R \Gamma
I}\right)}{\kappa_y^*\times OD_y}\frac{T_D}{2}.
\end{equation}
This equation is plotted as a solid line in Fig. \ref{TyODy}
using the results for $R$ and $\kappa_y$ (normalized to steady
state optical density of 4.5) from the fit to the data in Fig.
\ref{Tyint}. The agreement with the experimental data is
surprisingly good, since $\kappa_y$ and $\kappa$ are not strictly
linear in the optical density (see Eqns. \ref{kappar},
\ref{kappa}) and the heating rate depends on temperature and
density.

\section{Conclusion}
We have presented a general Doppler cooling technique which works
for optically dense, trapped atomic samples. In experiments with
magnetically trapped Chromium atoms we have achieved an increase
in phase space density by two orders of magnitude. To explain our
experimental results, in particular the fast and efficient
three-dimensional cooling in a one-dimensional molasses, we have
developed a model which takes into account radial cooling due to
reabsorption of scattered photons. Our experimental findings for
the intensity-dependence of the cooling rate and the steady-state
temperature agree well with the theoretical considerations. We
could exclude thermalization effects via elastic collisions or
anharmonic mixing as the origin of the radial cooling we observe.
Optimum radial cooling is observed in spherical trap
configurations and the radial cooling efficiency increases with
the radial optical density.

Our technique should be applicable to most atomic species which
can be laser-cooled provided that a moderate magnetic offset-field
(e.g. $B_0=15$~G for $^{87}$Rb) is applied to prevent
depolarization of the sample. Implementation of this technique in
a BEC experiment could simplify the initial laser cooling stages,
since Doppler cooling in the trap reduces the figure of merit for
loading of the trap to the number of atoms transferred into the
trap regardless of temperature. Due to the low light intensities
needed to cool the atoms, our method might also be applicable to
magnetically trapped atoms in a cryogenic environment
\cite{Weinstein:2002,Carvalho:2002}. In our experiment, the
achieved phase space density for chromium already provides good
starting conditions for evaporation and only excessive atom loss
and heating due to dipolar relaxation \cite{Schmidt:2002b} in the
consecutive evaporation stage prevented us from reaching quantum
degeneracy.

\begin{acknowledgments}
This work was funded by the Forschergruppe Quantengase der
Deutschen Forschungsgemeinschaft. P.O.S has been supported by the
Studien\-stiftung des deutschen Volkes.
\end{acknowledgments}


\begin{thebibliography}{29}
\expandafter\ifx\csname
natexlab\endcsname\relax\def\natexlab#1{#1}\fi
\expandafter\ifx\csname bibnamefont\endcsname\relax
  \def\bibnamefont#1{#1}\fi
\expandafter\ifx\csname bibfnamefont\endcsname\relax
  \def\bibfnamefont#1{#1}\fi
\expandafter\ifx\csname citenamefont\endcsname\relax
  \def\citenamefont#1{#1}\fi
\expandafter\ifx\csname url\endcsname\relax
  \def\url#1{\texttt{#1}}\fi
\expandafter\ifx\csname
urlprefix\endcsname\relax\def\urlprefix{URL }\fi
\providecommand{\bibinfo}[2]{#2}
\providecommand{\eprint}[2][]{\url{#2}}

\bibitem[{\citenamefont{Lett et~al.}(1989)\citenamefont{Lett, Phillips,
  Rolston, Tanner, Watts, and Westbrook}}]{Lett:89}
\bibinfo{author}{\bibfnamefont{P.~D.} \bibnamefont{Lett}},
  \bibinfo{author}{\bibfnamefont{W.~D.} \bibnamefont{Phillips}},
  \bibinfo{author}{\bibfnamefont{S.~L.} \bibnamefont{Rolston}},
  \bibinfo{author}{\bibfnamefont{C.~E.} \bibnamefont{Tanner}},
  \bibinfo{author}{\bibfnamefont{R.~N.} \bibnamefont{Watts}}, \bibnamefont{and}
  \bibinfo{author}{\bibfnamefont{C.~I.} \bibnamefont{Westbrook}},
  \bibinfo{journal}{J.~Opt.~Soc.~Am.~B} \textbf{\bibinfo{volume}{6}},
  \bibinfo{pages}{2084} (\bibinfo{year}{1989}).

\bibitem[{\citenamefont{Ketterle and van Druten}(1996)}]{Ketterle1996c}
\bibinfo{author}{\bibfnamefont{W.}~\bibnamefont{Ketterle}} \bibnamefont{and}
  \bibinfo{author}{\bibfnamefont{N.}~\bibnamefont{van Druten}},
  \bibinfo{journal}{Adv. At. Mol. Opt. Phys.} \textbf{\bibinfo{volume}{37}},
  \bibinfo{pages}{181} (\bibinfo{year}{1996}).

\bibitem[{\citenamefont{Anglin and Ketterle}(2002)}]{Anglin:2002}
\bibinfo{author}{\bibfnamefont{J.~R.} \bibnamefont{Anglin}} \bibnamefont{and}
  \bibinfo{author}{\bibfnamefont{W.}~\bibnamefont{Ketterle}},
  \bibinfo{journal}{Nature} \textbf{\bibinfo{volume}{416}},
  \bibinfo{pages}{211} (\bibinfo{year}{2002}).

\bibitem[{\citenamefont{Inguscio et~al.}(1998)\citenamefont{Inguscio,
  Stringari, and Wieman}}]{EFSummer:98}
\bibinfo{editor}{\bibfnamefont{M.}~\bibnamefont{Inguscio}},
  \bibinfo{editor}{\bibfnamefont{S.}~\bibnamefont{Stringari}},
  \bibnamefont{and} \bibinfo{editor}{\bibfnamefont{C.~E.}
  \bibnamefont{Wieman}}, eds., \emph{\bibinfo{title}{Bose-Einstein Condensation
  in Atomic Gases}} (\bibinfo{organization}{International School of Physics
  ``Enrico Fermi''}, \bibinfo{year}{1998}).

\bibitem[{\citenamefont{Cornell and Wieman}(2002)}]{Cornell:2002}
\bibinfo{author}{\bibfnamefont{E.~A.} \bibnamefont{Cornell}} \bibnamefont{and}
  \bibinfo{author}{\bibfnamefont{C.~E.} \bibnamefont{Wieman}},
  \bibinfo{journal}{Rev.~Mod.~Phys} \textbf{\bibinfo{volume}{74}}
  (\bibinfo{year}{2002}).

\bibitem[{\citenamefont{Dalibard and Cohen-Tannoudji}(1989)}]{Dalibard1989a}
\bibinfo{author}{\bibfnamefont{J.}~\bibnamefont{Dalibard}} \bibnamefont{and}
  \bibinfo{author}{\bibfnamefont{C.}~\bibnamefont{Cohen-Tannoudji}},
  \bibinfo{journal}{J. Opt. Soc. Am. B} \textbf{\bibinfo{volume}{6}},
  \bibinfo{pages}{2023} (\bibinfo{year}{1989}).

\bibitem[{\citenamefont{Barrett et~al.}(2001)\citenamefont{Barrett, Sauer, and
  Chapman}}]{Barrett2001a}
\bibinfo{author}{\bibfnamefont{M.}~\bibnamefont{Barrett}},
  \bibinfo{author}{\bibfnamefont{J.}~\bibnamefont{Sauer}}, \bibnamefont{and}
  \bibinfo{author}{\bibfnamefont{M.}~\bibnamefont{Chapman}},
  \bibinfo{journal}{Phys. Rev. Lett.} \textbf{\bibinfo{volume}{87}},
  \bibinfo{pages}{010404} (\bibinfo{year}{2001}).

\bibitem[{\citenamefont{Hijmans et~al.}(1989)\citenamefont{Hijmans, Luiten,
  Setija, and Walraven}}]{Hijmans1989a}
\bibinfo{author}{\bibfnamefont{T.}~\bibnamefont{Hijmans}},
  \bibinfo{author}{\bibfnamefont{O.}~\bibnamefont{Luiten}},
  \bibinfo{author}{\bibfnamefont{I.}~\bibnamefont{Setija}}, \bibnamefont{and}
  \bibinfo{author}{\bibfnamefont{J.~T.~M.} \bibnamefont{Walraven}},
  \bibinfo{journal}{J. Opt. Soc. Am. B} \textbf{\bibinfo{volume}{6}},
  \bibinfo{pages}{2235} (\bibinfo{year}{1989}).

\bibitem[{\citenamefont{Helmerson et~al.}(1992)\citenamefont{Helmerson, Martin,
  and Pritchard}}]{Helmerson1992a}
\bibinfo{author}{\bibfnamefont{K.}~\bibnamefont{Helmerson}},
  \bibinfo{author}{\bibfnamefont{A.}~\bibnamefont{Martin}}, \bibnamefont{and}
  \bibinfo{author}{\bibfnamefont{D.~E.} \bibnamefont{Pritchard}},
  \bibinfo{journal}{J. Opt. Soc. Am. B} \textbf{\bibinfo{volume}{9}},
  \bibinfo{pages}{1988} (\bibinfo{year}{1992}).

\bibitem[{\citenamefont{Setija et~al.}(1993)\citenamefont{Setija, Werij,
  Luiten, Reynolds, Hijmans, and Walraven}}]{Setija:93}
\bibinfo{author}{\bibfnamefont{I.~D.} \bibnamefont{Setija}},
  \bibinfo{author}{\bibfnamefont{H.}~\bibnamefont{Werij}},
  \bibinfo{author}{\bibfnamefont{O.~J.} \bibnamefont{Luiten}},
  \bibinfo{author}{\bibfnamefont{M.~W.} \bibnamefont{Reynolds}},
  \bibinfo{author}{\bibfnamefont{T.~W.} \bibnamefont{Hijmans}},
  \bibnamefont{and} \bibinfo{author}{\bibfnamefont{J.~T.~M.}
  \bibnamefont{Walraven}}, \bibinfo{journal}{Phys.~Rev.~Lett.}
  \textbf{\bibinfo{volume}{70}}, \bibinfo{pages}{2257} (\bibinfo{year}{1993}).

\bibitem[{\citenamefont{Schreck et~al.}(2001)\citenamefont{Schreck, Ferrari,
  Corwin, Cubizolles, Khaykovich, Mewes, and Salomon}}]{Schreck:2001}
\bibinfo{author}{\bibfnamefont{F.}~\bibnamefont{Schreck}},
  \bibinfo{author}{\bibfnamefont{G.}~\bibnamefont{Ferrari}},
  \bibinfo{author}{\bibfnamefont{K.~L.} \bibnamefont{Corwin}},
  \bibinfo{author}{\bibfnamefont{J.}~\bibnamefont{Cubizolles}},
  \bibinfo{author}{\bibfnamefont{L.}~\bibnamefont{Khaykovich}},
  \bibinfo{author}{\bibfnamefont{M.-O.} \bibnamefont{Mewes}}, \bibnamefont{and}
  \bibinfo{author}{\bibfnamefont{C.}~\bibnamefont{Salomon}},
  \bibinfo{journal}{Phys.~Rev.~A} \textbf{\bibinfo{volume}{64}},
  \bibinfo{pages}{011402} (\bibinfo{year}{2001}).

\bibitem[{\citenamefont{Sesko et~al.}(1991)\citenamefont{Sesko, Walker, and
  Wieman}}]{Sesko1991a}
\bibinfo{author}{\bibfnamefont{D.}~\bibnamefont{Sesko}},
  \bibinfo{author}{\bibfnamefont{T.}~\bibnamefont{Walker}}, \bibnamefont{and}
  \bibinfo{author}{\bibfnamefont{C.}~\bibnamefont{Wieman}},
  \bibinfo{journal}{J. Opt. Soc. Am. B} \textbf{\bibinfo{volume}{8}},
  \bibinfo{pages}{946} (\bibinfo{year}{1991}).

\bibitem[{\citenamefont{Hillenbrand et~al.}(1994)\citenamefont{Hillenbrand,
  Foot, and Burnett}}]{Hillenbrand:1994}
\bibinfo{author}{\bibfnamefont{G.}~\bibnamefont{Hillenbrand}},
  \bibinfo{author}{\bibfnamefont{C.~J.} \bibnamefont{Foot}}, \bibnamefont{and}
  \bibinfo{author}{\bibfnamefont{K.}~\bibnamefont{Burnett}},
  \bibinfo{journal}{Phys.~Rev.~A} \textbf{\bibinfo{volume}{50}},
  \bibinfo{pages}{1479} (\bibinfo{year}{1994}).

\bibitem[{\citenamefont{Ellinger and Cooper}(1997)}]{Ellinger:1997}
\bibinfo{author}{\bibfnamefont{K.}~\bibnamefont{Ellinger}} \bibnamefont{and}
  \bibinfo{author}{\bibfnamefont{J.}~\bibnamefont{Cooper}},
  \bibinfo{journal}{Phys.~Rev.~A} \textbf{\bibinfo{volume}{55}},
  \bibinfo{pages}{4351} (\bibinfo{year}{1997}).

\bibitem[{\citenamefont{Castin et~al.}(1998)\citenamefont{Castin, Cirac, and
  Lewenstein}}]{Castin1998b}
\bibinfo{author}{\bibfnamefont{Y.}~\bibnamefont{Castin}},
  \bibinfo{author}{\bibfnamefont{J.}~\bibnamefont{Cirac}}, \bibnamefont{and}
  \bibinfo{author}{\bibfnamefont{M.}~\bibnamefont{Lewenstein}},
  \bibinfo{journal}{Phys. Rev. Lett.} \textbf{\bibinfo{volume}{80}},
  \bibinfo{pages}{5305} (\bibinfo{year}{1998}).

\bibitem[{\citenamefont{Sukumar and Brink}(1997)}]{Sukumar1997a}
\bibinfo{author}{\bibfnamefont{C.}~\bibnamefont{Sukumar}} \bibnamefont{and}
  \bibinfo{author}{\bibfnamefont{D.}~\bibnamefont{Brink}},
  \bibinfo{journal}{Phys. Rev. A} \textbf{\bibinfo{volume}{56}},
  \bibinfo{pages}{2451} (\bibinfo{year}{1997}).

\bibitem[{Dop()}]{Doppler_Foot1}
\bibinfo{note}{In principle the same argument holds for a $J^{'}=J$ transition;
  for simplicity we will concentrate on the situation above}.

\bibitem[{\citenamefont{Stuhler}(2002)}]{Stuhler:2002}
\bibinfo{author}{\bibfnamefont{J.}~\bibnamefont{Stuhler}},
  \bibinfo{type}{Dissertation}, \bibinfo{school}{Universit{\"a}t Konstanz,
  Lehrstuhl J. Mlynek}, \bibinfo{address}{Ufo-Verlag, Allensbach}
  (\bibinfo{year}{2002}).

\bibitem[{\citenamefont{Cohen-Tannoudji
  et~al.}(1992)\citenamefont{Cohen-Tannoudji, Dupont-Roc, and
  Grynberg}}]{Cohen:92doppler}
\bibinfo{author}{\bibfnamefont{C.}~\bibnamefont{Cohen-Tannoudji}},
  \bibinfo{author}{\bibfnamefont{J.}~\bibnamefont{Dupont-Roc}},
  \bibnamefont{and} \bibinfo{author}{\bibfnamefont{G.}~\bibnamefont{Grynberg}},
  \emph{\bibinfo{title}{Atom-photon interactions}} (\bibinfo{publisher}{John
  Wiley $\&$ Sons}, \bibinfo{year}{1992}), \bibinfo{edition}{1st} ed.,
  \bibinfo{note}{page 93ff.}

\bibitem[{\citenamefont{Walker and Feng}(1994)}]{Walker:1994}
\bibinfo{author}{\bibfnamefont{T.}~\bibnamefont{Walker}} \bibnamefont{and}
  \bibinfo{author}{\bibfnamefont{P.}~\bibnamefont{Feng}},
  \bibinfo{journal}{Advances in Atomic, Molecular, and Optical Physics}
  \textbf{\bibinfo{volume}{34}}, \bibinfo{pages}{125} (\bibinfo{year}{1994}).

\bibitem[{\citenamefont{Bergeman et~al.}(1987)\citenamefont{Bergeman, Erez, and
  Metcalf}}]{Bergeman1987a}
\bibinfo{author}{\bibfnamefont{T.}~\bibnamefont{Bergeman}},
  \bibinfo{author}{\bibfnamefont{G.}~\bibnamefont{Erez}}, \bibnamefont{and}
  \bibinfo{author}{\bibfnamefont{H.}~\bibnamefont{Metcalf}},
  \bibinfo{journal}{Phys. Rev. A} \textbf{\bibinfo{volume}{35}},
  \bibinfo{pages}{1535} (\bibinfo{year}{1987}).

\bibitem[{\citenamefont{Mewes et~al.}(1996)\citenamefont{Mewes, Andrews, van
  Druten, Kurn, Durfee, and Ketterle}}]{Mewes1996a}
\bibinfo{author}{\bibfnamefont{M.-O.} \bibnamefont{Mewes}},
  \bibinfo{author}{\bibfnamefont{M.}~\bibnamefont{Andrews}},
  \bibinfo{author}{\bibfnamefont{N.}~\bibnamefont{van Druten}},
  \bibinfo{author}{\bibfnamefont{D.}~\bibnamefont{Kurn}},
  \bibinfo{author}{\bibfnamefont{D.}~\bibnamefont{Durfee}}, \bibnamefont{and}
  \bibinfo{author}{\bibfnamefont{W.}~\bibnamefont{Ketterle}},
  \bibinfo{journal}{Phys. Rev. Lett.} \textbf{\bibinfo{volume}{77}},
  \bibinfo{pages}{416} (\bibinfo{year}{1996}).

\bibitem[{\citenamefont{Stuhler et~al.}(2001)\citenamefont{Stuhler, Schmidt,
  Hensler, Werner, Mlynek, and Pfau}}]{Stuhler:2001}
\bibinfo{author}{\bibfnamefont{J.}~\bibnamefont{Stuhler}},
  \bibinfo{author}{\bibfnamefont{P.~O.} \bibnamefont{Schmidt}},
  \bibinfo{author}{\bibfnamefont{S.}~\bibnamefont{Hensler}},
  \bibinfo{author}{\bibfnamefont{J.}~\bibnamefont{Werner}},
  \bibinfo{author}{\bibfnamefont{J.}~\bibnamefont{Mlynek}}, \bibnamefont{and}
  \bibinfo{author}{\bibfnamefont{T.}~\bibnamefont{Pfau}},
  \bibinfo{journal}{Phys.~Rev.~A} \textbf{\bibinfo{volume}{64}},
  \bibinfo{pages}{031405} (\bibinfo{year}{2001}).

\bibitem[{\citenamefont{Schmidt and et~al.}({\natexlab{a}})}]{Schmidt:2002a}
\bibinfo{author}{\bibfnamefont{P.~O.} \bibnamefont{Schmidt}} \bibnamefont{and}
  \bibinfo{author}{\bibnamefont{et~al.}}, \emph{\bibinfo{title}{Continuous
  loading of a ioffe-pritchard trap}}, \bibinfo{howpublished}{in preparation}.

\bibitem[{\citenamefont{Bell et~al.}(1999)\citenamefont{Bell, Stuhler, Locher,
  Hensler, Mlynek, and Pfau}}]{Bell:99}
\bibinfo{author}{\bibfnamefont{A.~S.} \bibnamefont{Bell}},
  \bibinfo{author}{\bibfnamefont{J.}~\bibnamefont{Stuhler}},
  \bibinfo{author}{\bibfnamefont{S.}~\bibnamefont{Locher}},
  \bibinfo{author}{\bibfnamefont{S.}~\bibnamefont{Hensler}},
  \bibinfo{author}{\bibfnamefont{J.}~\bibnamefont{Mlynek}}, \bibnamefont{and}
  \bibinfo{author}{\bibfnamefont{T.}~\bibnamefont{Pfau}},
  \bibinfo{journal}{Europhys.~Lett.} \textbf{\bibinfo{volume}{45}},
  \bibinfo{pages}{156} (\bibinfo{year}{1999}).

\bibitem[{\citenamefont{Bradley et~al.}(2000)\citenamefont{Bradley, McClelland,
  Anderson, and Celotta}}]{Bradley:00a}
\bibinfo{author}{\bibfnamefont{C.~C.} \bibnamefont{Bradley}},
  \bibinfo{author}{\bibfnamefont{J.~J.} \bibnamefont{McClelland}},
  \bibinfo{author}{\bibfnamefont{W.~R.} \bibnamefont{Anderson}},
  \bibnamefont{and} \bibinfo{author}{\bibfnamefont{R.~J.}
  \bibnamefont{Celotta}}, \bibinfo{journal}{Phys.~Rev.~A}
  \textbf{\bibinfo{volume}{61}}, \bibinfo{pages}{053407}
  (\bibinfo{year}{2000}).

\bibitem[{\citenamefont{Schmidt and et~al.}({\natexlab{b}})}]{Schmidt:2002b}
\bibinfo{author}{\bibfnamefont{P.~O.} \bibnamefont{Schmidt}} \bibnamefont{and}
  \bibinfo{author}{\bibnamefont{et~al.}}, \emph{\bibinfo{title}{Dipolar
  relaxation in ultracold dipolar gases}}, \bibinfo{howpublished}{in
  preparation}.

\bibitem[{\citenamefont{Weinstein et~al.}(2002)\citenamefont{Weinstein,
  deCarvalho, Hancox, and Doyle}}]{Weinstein:2002}
\bibinfo{author}{\bibfnamefont{J.~D.} \bibnamefont{Weinstein}},
  \bibinfo{author}{\bibfnamefont{R.}~\bibnamefont{deCarvalho}},
  \bibinfo{author}{\bibfnamefont{C.~I.} \bibnamefont{Hancox}},
  \bibnamefont{and} \bibinfo{author}{\bibfnamefont{J.~M.} \bibnamefont{Doyle}},
  \bibinfo{journal}{Phys. Rev. A} \textbf{\bibinfo{volume}{65}},
  \bibinfo{pages}{021604} (\bibinfo{year}{2002}).

\bibitem[{\citenamefont{deCarvalho~et al.}()}]{Carvalho:2002}
\bibinfo{author}{\bibfnamefont{R.}~\bibnamefont{deCarvalho~et al.}},
  \bibinfo{howpublished}{same issue}.

\end{thebibliography}

\end{document}